\documentclass[letters,usenatbib]{mnras}
\usepackage[T1]{fontenc}
\DeclareRobustCommand{\VAN}[3]{#2}
\let\VANthebibliography\thebibliography
\def\thebibliography{\DeclareRobustCommand{\VAN}[3]{##3}\VANthebibliography}

\usepackage[dvipsnames]{xcolor} 
\usepackage{svg}
\usepackage{amsmath} 
\usepackage{amssymb}
\usepackage{booktabs}
\usepackage{bookmark}
\usepackage{hyperref}
\usepackage[utf8]{inputenc}
\usepackage{graphicx} 
\usepackage{lineno}
\usepackage[normalem]{ulem}
\usepackage{xspace}
\usepackage{orcidlink}

\newcommand{\ssout}[1]{}%{\sout{#1}}

\newcommand{\SSM}{SSM\xspace}
\newcommand{\DM}{DM\xspace}

\newcommand{\SNRLossFromFlow}{7\%}

\newcommand{\msun}{\ensuremath{M_\odot}\xspace}
\newcommand{\MinMatch}{\ensuremath{0.97}\xspace}

\newcommand{\MinPrimaryMass}{\ensuremath{0.2}\xspace}
\newcommand{\MinSecondaryMass}{\ensuremath{0.2}\xspace}
\newcommand{\MaxPrimaryMass}{\ensuremath{10 }\xspace}
\newcommand{\MaxSecondaryMass}{\ensuremath{1.0}\xspace}
\newcommand{\MaxLowMassComponentSpin}{\ensuremath{0.1}\xspace}
\newcommand{\MaxHighMassComponentSpin}{\ensuremath{0.9}\xspace}
\newcommand{\MinMassRatio}{\ensuremath{0.1}\xspace}

\newcommand{\FARa}{$0.2 \thinspace \rm{yr}^{-1}$\xspace}
\newcommand{\fPBHlim}{$0.6$\xspace}

\newcommand{\GstLALbestFAR}{$0.2 \thinspace \rm{yr}^{-1}$\xspace}

\newcommand{\MBTAbestFAR}{$1.4 \thinspace \rm{yr}^{-1}$\xspace}

\newcommand{\PyCBCbestFAR}{$0.14 \thinspace \rm{yr}^{-1}$\xspace}

\def\apjl{{\it Astrophysical Journal Letters}\xspace}

\def\mnras{{\it Monthly Notices of the RAS}\xspace}

\title[Search for SSM BBHs in aLIGO and AdV in O3]{Search for subsolar-mass black hole binaries in the second part of Advanced LIGO's and Advanced Virgo's third observing run}

\author[LVK]{%
R.~Abbott$^{1}$ , %rich.abbott
H.~Abe$^{2}$ , %homare.abe
F.~Acernese$^{3,4}$ , %fausto.acernese
K.~Ackley\,\orcidlink{0000-0002-8648-0767}$^{5}$ , %kendall.ackley
\newauthor
S.~Adhicary$^{6}$ , %shomik.adhicary
N.~Adhikari\,\orcidlink{0000-0002-4559-8427}$^{7}$ , %naresh.adhikari
R.~X.~Adhikari\,\orcidlink{0000-0002-5731-5076}$^{1}$ , %rana.adhikari
\newauthor
V.~K.~Adkins$^{8}$ , %vernita.adkins
V.~B.~Adya$^{9}$ , %vaishali.adya
C.~Affeldt$^{10,11}$
% ...
\newauthor
(The LIGO Scientific Collaboration, the Virgo Collaboration, and the KAGRA Collaboration)
\newauthor
and
\newauthor
S.~Shandera$^{6}$ and D.~Jeong$^{6}$
\\
${}^{\ast}$Deceased, December 2021.
}\date{
$^{1}$LIGO Laboratory, California Institute of Technology, Pasadena, CA 91125, USA % {CIT} {1}
\\
$^{2}$Graduate School of Science, Tokyo Institute of Technology, Meguro-ku, Tokyo 152-8551, Japan % {jp.TITECH} {2}
\\
$^{3}$Dipartimento di Farmacia, Universit\`a di Salerno, I-84084 Fisciano, Salerno, Italy % {DIFASA} {3}
\\
$^{4}$INFN, Sezione di Napoli, I-80126 Napoli, Italy % {INFNNA} {4}
\\
$^{5}$OzGrav, School of Physics \& Astronomy, Monash University, Clayton 3800, Victoria, Australia % {MonashU} {5}
\\
$^{6}$The Pennsylvania State University, University Park, PA 16802, USA % {PennState} {6}
\\
$^{7}$University of Wisconsin-Milwaukee, Milwaukee, WI 53201, USA % {UWM} {7}
\\
$^{8}$Louisiana State University, Baton Rouge, LA 70803, USA % {LSU} {8}
\\
$^{9}$OzGrav, Australian National University, Canberra, Australian Capital Territory 0200, Australia % {ANU} {9}
\\
$^{10}$Max Planck Institute for Gravitational Physics (Albert Einstein Institute), D-30167 Hannover, Germany % {AEIHannover} {10}
\\
$^{11}$Leibniz Universit\"at Hannover, D-30167 Hannover, Germany % {Leibniz} {11}
\\
$^{12}$Inter-University Centre for Astronomy and Astrophysics, Pune 411007, India % {IUCAA} {12}
\\
$^{13}$University of Cambridge, Cambridge CB2 1TN, United Kingdom % {Cambridge} {13}
\\
$^{14}$Theoretisch-Physikalisches Institut, Friedrich-Schiller-Universit\"at Jena, D-07743 Jena, Germany % {UNIJENA} {14}
\\
$^{15}$Instituto Nacional de Pesquisas Espaciais, 12227-010 S\~{a}o Jos\'{e} dos Campos, S\~{a}o Paulo, Brazil % {GWINPE} {15}
\\
$^{16}$Cardiff University, Cardiff CF24 3AA, United Kingdom % {Cardiff} {16}
\\
$^{17}$INFN, Sezione di Pisa, I-56127 Pisa, Italy % {INFNPI} {17}
\\
$^{18}$International Centre for Theoretical Sciences, Tata Institute of Fundamental Research, Bengaluru 560089, India % {ICTS-TIFR} {18}
\\
$^{19}$Gravitational Wave Science Project, National Astronomical Observatory of Japan (NAOJ), Mitaka City, Tokyo 181-8588, Japan % {jp.NAOJ.GW} {19}
\\
$^{20}$Advanced Technology Center, National Astronomical Observatory of Japan (NAOJ), Mitaka City, Tokyo 181-8588, Japan % {jp.NAOJ.ATC} {20}
\\
$^{21}$Dipartimento di Fisica, Universit\`a degli Studi di Torino, I-10125 Torino, Italy % {UNITO} {21}
\\
$^{22}$INFN Sezione di Torino, I-10125 Torino, Italy % {INFNTO} {22}
\\
$^{23}$SUPA, University of Glasgow, Glasgow G12 8QQ, United Kingdom % {Glasgow} {23}
\\
$^{24}$Univ. Savoie Mont Blanc, CNRS, Laboratoire d'Annecy de Physique des Particules - IN2P3, F-74000 Annecy, France % {LAPP} {24}
\\
$^{25}$Universit\`a di Napoli ``Federico II'', I-80126 Napoli, Italy % {UNINA} {25}
\\
$^{26}$Maastricht University, 6200 MD Maastricht, Netherlands % {UNIMAAST} {26}
\\
$^{27}$Nikhef, 1098 XG Amsterdam, Netherlands % {NIKHEF} {27}
\\
$^{28}$Department of Physics, The University of Tokyo, Bunkyo-ku, Tokyo 113-0033, Japan % {jp.UT.PHYS} {28}
\\
$^{29}$Research Center for the Early Universe (RESCEU), The University of Tokyo, Bunkyo-ku, Tokyo 113-0033, Japan % {jp.RESCEU} {29}
\\
$^{30}$Institut de Ci\`encies del Cosmos (ICCUB), Universitat de Barcelona, Barcelona, 08028, Spain % {ICCUB} {30}
\\
$^{31}$Institut de F\'{\i}sica d'Altes Energies (IFAE), Barcelona Institute of Science and Technology, and ICREA, E-08193 Barcelona, Spain % {IFAE} {31}
\\
$^{32}$Gran Sasso Science Institute (GSSI), I-67100 L'Aquila, Italy % {GSSI} {32}
\\
$^{33}$Dipartimento di Scienze Matematiche, Informatiche e Fisiche, Universit\`a di Udine, I-33100 Udine, Italy % {DIMAUD} {33}
\\
$^{34}$INFN, Sezione di Trieste, I-34127 Trieste, Italy % {INFNTS} {34}
\\
$^{35}$Embry-Riddle Aeronautical University, Prescott, AZ 86301, USA % {EmbryRiddle} {35}
\\
$^{36}$Artemis, Universit\'e C\^ote d'Azur, Observatoire de la C\^ote d'Azur, CNRS, F-06304 Nice, France % {ARTEMIS} {36}
\\
$^{37}$GRAPPA, Anton Pannekoek Institute for Astronomy and Institute for High-Energy Physics, University of Amsterdam, 1098 XH Amsterdam, Netherlands % {GRAPPA} {37}
\\
$^{38}$Department of Physics, National and Kapodistrian University of Athens, 15771 Ilissia, Greece % {UNIATHENS} {38}
\\
$^{39}$INFN, Sezione di Perugia, I-06123 Perugia, Italy % {INFNPG} {39}
\\
$^{40}$Universit\`a di Camerino, Dipartimento di Fisica, I-62032 Camerino, Italy % {UNICAM} {40}
\\
$^{41}$American University, Washington, D.C. 20016, USA % {American} {41}
\\
$^{42}$Earthquake Research Institute, The University of Tokyo, Bunkyo-ku, Tokyo 113-0032, Japan % {jp.UT.ERI} {42}
\\
$^{43}$California State University Fullerton, Fullerton, CA 92831, USA % {Fullerton} {43}
\\
$^{44}$Universit\'e de Paris, CNRS, Astroparticule et Cosmologie, F-75006 Paris, France % {APC} {44}
\\
$^{45}$Universit\'e Paris-Saclay, CNRS/IN2P3, IJCLab, 91405 Orsay, France % {IJCLAB} {45}
\\
$^{46}$European Gravitational Observatory (EGO), I-56021 Cascina, Pisa, Italy % {EGO} {46}
\\
$^{47}$Georgia Institute of Technology, Atlanta, GA 30332, USA % {GaTech} {47}
\\
$^{48}$Chennai Mathematical Institute, Chennai 603103, India % {CMI} {48}
\\
$^{49}$Department of Mathematics and Physics, Graduate School of Science and Technology, Hirosaki University, Hirosaki, Aomori 036-8561, Japan % {jp.HIROSAKI} {49}
\\
$^{50}$Royal Holloway, University of London, London TW20 0EX, United Kingdom % {RHUL} {50}
\\
$^{51}$Kamioka Branch, National Astronomical Observatory of Japan (NAOJ), Kamioka-cho, Hida City, Gifu 506-1205, Japan % {jp.NAOJ.MOZ} {51}
\\
$^{52}$The Graduate University for Advanced Studies (SOKENDAI), Mitaka City, Tokyo 181-8588, Japan % {jp.SOKEN.NAOJ} {52}
\\
$^{53}$Universit\`a degli Studi di Urbino ``Carlo Bo'', I-61029 Urbino, Italy % {UNIURB} {53}
\\
$^{54}$INFN, Sezione di Firenze, I-50019 Sesto Fiorentino, Firenze, Italy % {INFNFI} {54}
\\
$^{55}$LIGO Livingston Observatory, Livingston, LA 70754, USA % {LLO} {55}
\\
$^{56}$INFN, Sezione di Roma, I-00185 Roma, Italy % {INFNRM} {56}
\\
$^{57}$Universit\'e catholique de Louvain, B-1348 Louvain-la-Neuve, Belgium % {UNILOUVAIN} {57}
\\
$^{58}$King's College London, University of London, London WC2R 2LS, United Kingdom % {KCL} {58}
\\
$^{59}$Korea Institute of Science and Technology Information, Daejeon 34141, Republic of Korea % {KISTI} {59}
\\
$^{60}$National Institute for Mathematical Sciences, Daejeon 34047, Republic of Korea % {NIMS} {60}
\\
$^{61}$Kenyon College, Gambier, OH 43022, USA % {Kenyon} {61}
\\
$^{62}$School of High Energy Accelerator Science, The Graduate University for Advanced Studies (SOKENDAI), Tsukuba City, Ibaraki 305-0801, Japan % {jp.SOKEN.KEK} {62}
\\
$^{63}$Institute for Gravitational and Subatomic Physics (GRASP), Utrecht University, 3584 CC Utrecht, Netherlands % {GRASP} {63}
\\
$^{64}$University of Oregon, Eugene, OR 97403, USA % {UOregon} {64}
\\
$^{65}$Syracuse University, Syracuse, NY 13244, USA % {Syracuse} {65}
\\
$^{66}$Universit\'e de Li\`ege, B-4000 Li\`ege, Belgium % {UNILIEGE} {66}
\\
$^{67}$Northwestern University, Evanston, IL 60208, USA % {Northwestern} {67}
\\
$^{68}$LIGO Hanford Observatory, Richland, WA 99352, USA % {LHO} {68}
\\
$^{69}$Dipartimento di Medicina, Chirurgia e Odontoiatria ``Scuola Medica Salernitana'', Universit\`a di Salerno, I-84081 Baronissi, Salerno, Italy % {DIMESA} {69}
\\
$^{70}$LIGO Laboratory, Massachusetts Institute of Technology, Cambridge, MA 02139, USA % {MIT} {70}
\\
$^{71}$Wigner RCP, RMKI, H-1121 Budapest, Hungary % {WIGNER} {71}
\\
$^{72}$University of Florida, Gainesville, FL 32611, USA % {UFlorida} {72}
\\
$^{73}$Stanford University, Stanford, CA 94305, USA % {Stanford} {73}
\\
$^{74}$Universit\`a di Pisa, I-56127 Pisa, Italy % {UNIPI} {74}
\\
$^{75}$Universit\`a di Perugia, I-06123 Perugia, Italy % {UNIPG} {75}
\\
$^{76}$Universit\`a di Padova, Dipartimento di Fisica e Astronomia, I-35131 Padova, Italy % {UNIPD} {76}
\\
$^{77}$INFN, Sezione di Padova, I-35131 Padova, Italy % {INFNPD} {77}
\\
$^{78}$Montana State University, Bozeman, MT 59717, USA % {MontanaState} {78}
\\
$^{79}$Institute for Plasma Research, Bhat, Gandhinagar 382428, India % {IPR-Bhat} {79}
\\
$^{80}$Universiteit Gent, B-9000 Gent, Belgium % {UNIGENT} {80}
\\
$^{81}$Nicolaus Copernicus Astronomical Center, Polish Academy of Sciences, 00-716, Warsaw, Poland % {COPERNICUS} {81}
\\
$^{82}$Dipartimento di Ingegneria, Universit\`a del Sannio, I-82100 Benevento, Italy % {DINGSA} {82}
\\
$^{83}$OzGrav, University of Adelaide, Adelaide, South Australia 5005, Australia % {UAdelaide} {83}
\\
$^{84}$University of Minnesota, Minneapolis, MN 55455, USA % {UMinnesota} {84}
\\
$^{85}$Universit\"at Hamburg, D-22761 Hamburg, Germany % {ILP-UH} {85}
\\
$^{86}$SUPA, University of Strathclyde, Glasgow G1 1XQ, United Kingdom % {Strathclyde} {86}
\\
$^{87}$IAC3--IEEC, Universitat de les Illes Balears, E-07122 Palma de Mallorca, Spain % {BalearicIslands} {87}
\\
$^{88}$Departamento de Matem\'aticas, Universitat Aut\`onoma de Barcelona, 08193 Bellaterra (Barcelona), Spain % {DEMABAR} {88}
\\
$^{89}$INFN, Sezione di Genova, I-16146 Genova, Italy % {INFNGE} {89}
\\
$^{90}$OzGrav, University of Western Australia, Crawley, Western Australia 6009, Australia % {UWesternAustralia} {90}
\\
$^{91}$RRCAT, Indore, Madhya Pradesh 452013, India % {RRCAT} {91}
\\
$^{92}$Missouri University of Science and Technology, Rolla, MO 65409, USA % {MST} {92}
\\
$^{93}$Department of Physics and Astronomy, Vrije Universiteit Amsterdam, 1081 HV Amsterdam, Netherlands % {UNIVRIJE} {93}
\\
$^{94}$Lomonosov Moscow State University, Moscow 119991, Russia % {MoscowState} {94}
\\
$^{95}$Center for Theoretical Physics, Polish Academy of Sciences, 02-668, Warsaw, Poland % {CTPWAR} {95}
\\
$^{96}$Universit\`a di Trento, Dipartimento di Fisica, I-38123 Povo, Trento, Italy % {UNITN} {96}
\\
$^{97}$INFN, Trento Institute for Fundamental Physics and Applications, I-38123 Povo, Trento, Italy % {TIFPA} {97}
\\
$^{98}$Bar-Ilan University, Ramat Gan, 5290002, Israel % {BarIlan} {98}
\\
$^{99}$Dipartimento di Fisica ``E.R. Caianiello'', Universit\`a di Salerno, I-84084 Fisciano, Salerno, Italy % {DIFISA} {99}
\\
$^{100}$INFN, Sezione di Napoli, Gruppo Collegato di Salerno, I-80126 Napoli, Italy % {INFNSA} {100}
\\
$^{101}$Universit\`a di Roma ``La Sapienza'', I-00185 Roma, Italy % {UNIRM} {101}
\\
$^{102}$Univ Rennes, CNRS, Institut FOTON - UMR 6082, F-3500 Rennes, France % {UNIRENN} {102}
\\
$^{103}$Indian Institute of Technology Bombay, Powai, Mumbai 400 076, India % {IIT-Bombay} {103}
\\
$^{104}$INFN, Laboratori Nazionali del Gran Sasso, I-67100 Assergi, Italy % {LNGS} {104}
\\
$^{105}$Laboratoire Kastler Brossel, Sorbonne Universit\'e, CNRS, ENS-Universit\'e PSL, Coll\`ege de France, F-75005 Paris, France % {LKB} {105}
\\
$^{106}$Christopher Newport University, Newport News, VA 23606, USA % {CNU} {106}
\\
$^{107}$University of Birmingham, Birmingham B15 2TT, United Kingdom % {Birmingham} {107}
\\
$^{108}$Astronomical Observatory Warsaw University, 00-478 Warsaw, Poland % {AOWAR} {108}
\\
$^{109}$University of Maryland, College Park, MD 20742, USA % {UMaryland} {109}
\\
$^{110}$Max Planck Institute for Gravitational Physics (Albert Einstein Institute), D-14476 Potsdam, Germany % {AEIPotsdam} {110}
\\
$^{111}$Universit\`a degli Studi di Milano-Bicocca, I-20126 Milano, Italy % {UNIMIB} {111}
\\
$^{112}$INFN, Sezione di Milano-Bicocca, I-20126 Milano, Italy % {INFNMIB} {112}
\\
$^{113}$L2IT, Laboratoire des 2 Infinis - Toulouse, Universit\'e de Toulouse, CNRS/IN2P3, UPS, F-31062 Toulouse Cedex 9, France % {L2IT} {113}
\\
$^{114}$University of Portsmouth, Portsmouth, PO1 3FX, United Kingdom % {Portsmouth} {114}
\\
$^{115}$Universit\'e de Lyon, Universit\'e Claude Bernard Lyon 1, CNRS, Institut Lumi\`ere Mati\`ere, F-69622 Villeurbanne, France % {ILMLYON} {115}
\\
$^{116}$IGFAE, Universidade de Santiago de Compostela, 15782 Spain % {USDC} {116}
\\
$^{117}$Stony Brook University, Stony Brook, NY 11794, USA % {SBU} {117}
\\
$^{118}$Center for Computational Astrophysics, Flatiron Institute, New York, NY 10010, USA % {CCA} {118}
\\
$^{119}$NASA Goddard Space Flight Center, Greenbelt, MD 20771, USA % {Goddard} {119}
\\
$^{120}$Dipartimento di Fisica, Universit\`a degli Studi di Genova, I-16146 Genova, Italy % {UNIGE} {120}
\\
$^{121}$Department of Astronomy, Beijing Normal University, Beijing 100875, China % {cn.BNU} {121}
\\
$^{122}$Texas A\&M University, College Station, TX 77843, USA % {TAMU} {122}
\\
$^{123}$OzGrav, University of Melbourne, Parkville, Victoria 3010, Australia % {UMelbourne} {123}
\\
$^{124}$Universit\`a degli Studi di Sassari, I-07100 Sassari, Italy % {UNISS} {124}
\\
$^{125}$INFN, Laboratori Nazionali del Sud, I-95125 Catania, Italy % {LNS} {125}
\\
$^{126}$Universit\`a di Roma Tor Vergata, I-00133 Roma, Italy % {UNITOV} {126}
\\
$^{127}$INFN, Sezione di Roma Tor Vergata, I-00133 Roma, Italy % {INFNTOV} {127}
\\
$^{128}$University of Sannio at Benevento, I-82100 Benevento, Italy and INFN, Sezione di Napoli, I-80100 Napoli, Italy % {USannio} {128}
\\
$^{129}$Departamento de Astronom\'{\i}a y Astrof\'{\i}sica, Universitat de Val\`encia, E-46100 Burjassot, Val\`encia, Spain % {DEASVAL} {129}
\\
$^{130}$Rochester Institute of Technology, Rochester, NY 14623, USA % {RIT} {130}
\\
$^{131}$National Tsing Hua University, Hsinchu City, 30013 Taiwan, Republic of China % {NTHU} {131}
\\
$^{132}$The Chinese University of Hong Kong, Shatin, NT, Hong Kong % {CUHK} {132}
\\
$^{133}$Department of Applied Physics, Fukuoka University, Jonan, Fukuoka City, Fukuoka 814-0180, Japan % {jp.FUKUOKA} {133}
\\
$^{134}$OzGrav, Charles Sturt University, Wagga Wagga, New South Wales 2678, Australia % {CharlesSturt} {134}
\\
$^{135}$Department of Physics, Tamkang University, Danshui Dist., New Taipei City 25137, Taiwan % {tw.TAMKANG} {135}
\\
$^{136}$Department of Physics, Center for High Energy and High Field Physics, National Central University, Zhongli District, Taoyuan City 32001, Taiwan % {tw.NCU} {136}
\\
$^{137}$CaRT, California Institute of Technology, Pasadena, CA 91125, USA % {CaRT} {137}
\\
$^{138}$Dipartimento di Ingegneria Industriale (DIIN), Universit\`a di Salerno, I-84084 Fisciano, Salerno, Italy % {DIINSA} {138}
\\
$^{139}$Institute of Physics, Academia Sinica, Nankang, Taipei 11529, Taiwan % {tw.ACASINICA} {139}
\\
$^{140}$Universit\'e Lyon, Universit\'e Claude Bernard Lyon 1, CNRS, IP2I Lyon / IN2P3, UMR 5822, F-69622 Villeurbanne, France % {UNILYON} {140}
\\
$^{141}$INAF, Osservatorio Astronomico di Padova, I-35122 Padova, Italy % {INAFPD} {141}
\\
$^{142}$OzGrav, Swinburne University of Technology, Hawthorn VIC 3122, Australia % {Swinburne} {142}
\\
$^{143}$Universit\'e libre de Bruxelles, 1050 Bruxelles, Belgium % {UNIBRUX} {143}
\\
$^{144}$Universit\'{e} Libre de Bruxelles, Brussels 1050, Belgium % {ULB} {144}
\\
$^{145}$Departamento de Matem\'aticas, Universitat de Val\`encia, E-46100 Burjassot, Val\`encia, Spain % {DEMAVAL} {145}
\\
$^{146}$Texas Tech University, Lubbock, TX 79409, USA % {TTU} {146}
\\
$^{147}$University of British Columbia, Vancouver, BC V6T 1Z4, Canada % {UBC} {147}
\\
$^{148}$Columbia University, New York, NY 10027, USA % {Columbia} {148}
\\
$^{149}$University of Rhode Island, Kingston, RI 02881, USA % {URI} {149}
\\
$^{150}$The University of Texas Rio Grande Valley, Brownsville, TX 78520, USA % {CGWA-UTRGV} {150}
\\
$^{151}$Bellevue College, Bellevue, WA 98007, USA % {Bellevue} {151}
\\
$^{152}$Scuola Normale Superiore, I-56126 Pisa, Italy % {SNS} {152}
\\
$^{153}$E\"otv\"os University, Budapest 1117, Hungary % {Eotvos} {153}
\\
$^{154}$Villanova University, Villanova, PA 19085, USA % {Villanova} {154}
\\
$^{155}$The University of Sheffield, Sheffield S10 2TN, United Kingdom % {USheffield} {155}
\\
$^{156}$Universit\'e Lyon, Universit\'e Claude Bernard Lyon 1, CNRS, Laboratoire des Mat\'eriaux Avanc\'es (LMA), IP2I Lyon / IN2P3, UMR 5822, F-69622 Villeurbanne, France % {LMA} {156}
\\
$^{157}$Dipartimento di Scienze Matematiche, Fisiche e Informatiche, Universit\`a di Parma, I-43124 Parma, Italy % {UNIPR} {157}
\\
$^{158}$INFN, Sezione di Milano Bicocca, Gruppo Collegato di Parma, I-43124 Parma, Italy % {INFNPR} {158}
\\
$^{159}$The University of Utah, Salt Lake City, UT 84112, USA % {UUtah} {159}
\\
$^{160}$Carleton College, Northfield, MN 55057, USA % {Carleton} {160}
\\
$^{161}$National Center for Nuclear Research, 05-400 {\' S}wierk-Otwock, Poland % {NCNRPL} {161}
\\
$^{162}$Institut d’Astrophysique de Paris, Sorbonne Universit\'e, CNRS, UMR 7095, 75014 Paris, France % {SORBONNE} {162}
\\
$^{163}$University of Zurich, Winterthurerstrasse 190, 8057 Zurich, Switzerland % {UZH} {163}
\\
$^{164}$Perimeter Institute, Waterloo, ON N2L 2Y5, Canada % {Perimeter} {164}
\\
$^{165}$Universit\'e de Strasbourg, CNRS, IPHC UMR 7178, F-67000 Strasbourg, France % {UNISTRAS} {165}
\\
$^{166}$University of Chicago, Chicago, IL 60637, USA % {UChicago} {166}
\\
$^{167}$Montclair State University, Montclair, NJ 07043, USA % {MontclairState} {167}
\\
$^{168}$Colorado State University, Fort Collins, CO 80523, USA % {CSU} {168}
\\
$^{169}$Institute for Nuclear Research, H-4026 Debrecen, Hungary % {INRNDEB} {169}
\\
$^{170}$University of Texas, Austin, TX 78712, USA % {UTAustin} {170}
\\
$^{171}$CNR-SPIN, I-84084 Fisciano, Salerno, Italy % {CNRSA} {171}
\\
$^{172}$Scuola di Ingegneria, Universit\`a della Basilicata, I-85100 Potenza, Italy % {UNIBAS} {172}
\\
$^{173}$Observatori Astron\`omic, Universitat de Val\`encia, E-46980 Paterna, Val\`encia, Spain % {OAVAL} {173}
\\
$^{174}$Centro de F\'{\i}sica das Universidades do Minho e do Porto, Universidade do Minho, PT-4710-057 Braga, Portugal % {UNIMINHO} {174}
\\
$^{175}$Department of Astronomy, The University of Tokyo, Mitaka City, Tokyo 181-8588, Japan % {jp.UT.ASTRO} {175}
\\
$^{176}$Faculty of Engineering, Niigata University, Nishi-ku, Niigata City, Niigata 950-2181, Japan % {jp.NIIGATA.ENG} {176}
\\
$^{177}$Department of Physics, Graduate School of Science, Osaka City University, Sumiyoshi-ku, Osaka City, Osaka 558-8585, Japan % {jp.OCU} {177}
\\
$^{178}$Vanderbilt University, Nashville, TN 37235, USA % {Vanderbilt} {178}
\\
$^{179}$State Key Laboratory of Magnetic Resonance and Atomic and Molecular Physics, Innovation Academy for Precision Measurement Science and Technology (APM), Chinese Academy of Sciences, Xiao Hong Shan, Wuhan 430071, China % {cn.CAS.WUHAN.APS} {179}
\\
$^{180}$ % {es.UAMCSIC} {180}
\\
$^{181}$SUPA, University of the West of Scotland, Paisley PA1 2BE, United Kingdom % {UWS} {181}
\\
$^{182}$University of Szeged, D\'{o}m t\'{e}r 9, Szeged 6720, Hungary % {SZTE} {182}
\\
$^{183}$INAF, Osservatorio Astronomico di Capodimonte, I-80131 Napoli, Italy % {INAFNA} {183}
\\
$^{184}$Queen Mary University of London, London E1 4NS, United Kingdom % {QMUL} {184}
\\
$^{185}$Universit\'e de Normandie, ENSICAEN, UNICAEN, CNRS/IN2P3, LPC Caen, F-14000 Caen, France % {UNICAEN} {185}
\\
$^{186}$The University of Mississippi, University, MS 38677, USA % {UMiss} {186}
\\
$^{187}$University of Michigan, Ann Arbor, MI 48109, USA % {UMichigan} {187}
\\
$^{188}$Ulsan National Institute of Science and Technology, Ulsan 44919, Republic of Korea % {UNIST} {188}
\\
$^{189}$Shanghai Astronomical Observatory, Chinese Academy of Sciences, Shanghai 200030, China % {cn.CAS.SHANGHAI} {189}
\\
$^{190}$University of Tokyo, Tokyo, 113-0033, Japan. % {UTokyo} {190}
\\
$^{191}$Institute for Cosmic Ray Research (ICRR), KAGRA Observatory, The University of Tokyo, Kashiwa City, Chiba 277-8582, Japan % {jp.ICRR.KAGRA} {191}
\\
$^{192}$Faculty of Science, University of Toyama, Toyama City, Toyama 930-8555, Japan % {jp.TOYAMA.DSCI} {192}
\\
$^{193}$Institute for Cosmic Ray Research (ICRR), KAGRA Observatory, The University of Tokyo, Kamioka-cho, Hida City, Gifu 506-1205, Japan % {jp.ICRR.MOZ} {193}
\\
$^{194}$University of California, Berkeley, CA 94720, USA % {UCBerkeley} {194}
\\
$^{195}$California State University, Los Angeles, Los Angeles, CA 90032, USA % {CalStateLA} {195}
\\
$^{196}$Lancaster University, Lancaster LA1 4YW, United Kingdom % {Lancaster} {196}
\\
$^{197}$College of Industrial Technology, Nihon University, Narashino City, Chiba 275-8575, Japan % {jp.NIHON} {197}
\\
$^{198}$Rutherford Appleton Laboratory, Didcot OX11 0DE, United Kingdom % {RAL} {198}
\\
$^{199}$Department of Astronomy \& Space Science, Chungnam National University, Yuseong-gu, Daejeon 34134, Republic of Korea % {kr.CHUNGNAMN} {199}
\\
$^{200}$Department of Physical Sciences, Aoyama Gakuin University, Sagamihara City, Kanagawa 252-5258, Japan % {jp.AOYAMA} {200}
\\
$^{201}$Kavli Institute for Astronomy and Astrophysics, Peking University, Haidian District, Beijing 100871, China % {cn.PEKING} {201}
\\
$^{202}$Department of Physics, Aristotle University of Thessaloniki, 54124 Thessaloniki, Greece % {ARISTOT} {202}
\\
$^{203}$Graduate School of Science and Engineering, University of Toyama, Toyama City, Toyama 930-8555, Japan % {jp.TOYAMA.GSSE} {203}
\\
$^{204}$Nambu Yoichiro Institute of Theoretical and Experimental Physics (NITEP), Osaka City University, Sumiyoshi-ku, Osaka City, Osaka 558-8585, Japan % {jp.OCU.NYIT} {204}
\\
$^{205}$Directorate of Construction, Services \& Estate Management, Mumbai 400094, India % {DCSEM} {205}
\\
$^{206}$Universiteit Antwerpen, 2000 Antwerpen, Belgium % {UNIANTW} {206}
\\
$^{207}$University of Bia{\l}ystok, 15-424 Bia{\l}ystok, Poland % {UNIBIALI} {207}
\\
$^{208}$Ewha Womans University, Seoul 03760, Republic of Korea % {Ewha} {208}
\\
$^{209}$National Astronomical Observatories, Chinese Academic of Sciences, Chaoyang District, Beijing, China % {cn.CAS.NAOC} {209}
\\
$^{210}$School of Astronomy and Space Science, University of Chinese Academy of Sciences, Chaoyang District, Beijing, China % {cn.UCAS} {210}
\\
$^{211}$University of Southampton, Southampton SO17 1BJ, United Kingdom % {Southampton} {211}
\\
$^{212}$Institute for Cosmic Ray Research (ICRR), The University of Tokyo, Kashiwa City, Chiba 277-8582, Japan % {jp.ICRR} {212}
\\
$^{213}$Faculty of Science, University of Toyama, Toyama City, Toyama 930-8555, Japan % {jp.TOYAMA.SCI} {213}
\\
$^{214}$Institute for High-Energy Physics, University of Amsterdam, 1098 XH Amsterdam, Netherlands % {IHEPAMST} {214}
\\
$^{215}$ % {CAU} {215}
\\
$^{216}$University of Washington Bothell, Bothell, WA 98011, USA % {UWB} {216}
\\
$^{217}$Institute of Applied Physics, Nizhny Novgorod, 603950, Russia % {NizhnyNovgorod} {217}
\\
$^{218}$Inje University Gimhae, South Gyeongsang 50834, Republic of Korea % {InjeU} {218}
\\
$^{219}$Department of Physics, Myongji University, Yongin 17058, Republic of Korea % {kr.MYONGJI} {219}
\\
$^{220}$Sungkyunkwan University, Seoul 03063, Republic of Korea % {SKKU} {220}
\\
$^{221}$Bard College, Annandale-On-Hudson, NY 12504, USA % {Bard} {221}
\\
$^{222}$Institute of Particle and Nuclear Studies (IPNS), High Energy Accelerator Research Organization (KEK), Tsukuba City, Ibaraki 305-0801, Japan % {jp.KEK.IPNS} {222}
\\
$^{223}$Institute of Mathematics, Polish Academy of Sciences, 00656 Warsaw, Poland % {IMAPOL} {223}
\\
$^{224}$Instituto de Fisica Teorica, 28049 Madrid, Spain % {es.IFT} {224}
\\
$^{225}$Department of Physics, Nagoya University, Chikusa-ku, Nagoya, Aichi 464-8602, Japan % {jp.NAGOYA} {225}
\\
$^{226}$Universit\'{e} de Montr\'{e}al/Polytechnique, Montreal, Quebec H3T 1J4, Canada % {UMontreal} {226}
\\
$^{227}$Laboratoire Lagrange, Universit\'e C\^ote d'Azur, Observatoire C\^ote d'Azur, CNRS, F-06304 Nice, France % {LAGRANGE} {227}
\\
$^{228}$Seoul National University, Seoul 08826, Republic of Korea % {SeoulNationalU} {228}
\\
$^{229}$NAVIER, \'{E}cole des Ponts, Univ Gustave Eiffel, CNRS, Marne-la-Vall\'{e}e, France % {NAVIER} {229}
\\
$^{230}$Universit\`a di Firenze, Sesto Fiorentino I-50019, Italy % {UNIFI} {230}
\\
$^{231}$Department of Physics, National Cheng Kung University, Tainan City 701, Taiwan % {tw.NCKU} {231}
\\
$^{232}$School of Physics and Technology, Wuhan University, Wuhan, Hubei, 430072, China % {cn.WUHAN} {232}
\\
$^{233}$National Center for High-performance computing, National Applied Research Laboratories, Hsinchu Science Park, Hsinchu City 30076, Taiwan % {tw.NARL} {233}
\\
$^{234}$Department of Physics, National Taiwan Normal University, sec. 4, Taipei 116, Taiwan % {tw.NTNU} {234}
\\
$^{235}$NASA Marshall Space Flight Center, Huntsville, AL 35811, USA % {NASA-MSFC} {235}
\\
$^{236}$INFN, Sezione di Roma Tre, I-00146 Roma, Italy % {INFNRM3} {236}
\\
$^{237}$ESPCI, CNRS, F-75005 Paris, France % {ESPCI} {237}
\\
$^{238}$West Virginia University, Morgantown, WV 26506, USA % {WVU} {238}
\\
$^{239}$School of Physics Science and Engineering, Tongji University, Shanghai 200092, China % {UNISHANG} {239}
\\
$^{240}$ % {Tsinghua} {240}
\\
$^{241}$Dipartimento di Fisica, Universit\`a di Trieste, I-34127 Trieste, Italy % {DIFITS} {241}
\\
$^{242}$Institute for Photon Science and Technology, The University of Tokyo, Bunkyo-ku, Tokyo 113-8656, Japan % {jp.UT.IPST} {242}
\\
$^{243}$Indian Institute of Technology Madras, Chennai 600036, India % {IIT-Madras} {243}
\\
$^{244}$Institute of Space and Astronautical Science (JAXA), Chuo-ku, Sagamihara City, Kanagawa 252-0222, Japan % {jp.JAXA.ISAS} {244}
\\
$^{245}$Institut des Hautes Etudes Scientifiques, F-91440 Bures-sur-Yvette, France % {IHES} {245}
\\
$^{246}$Faculty of Law, Ryukoku University, Fushimi-ku, Kyoto City, Kyoto 612-8577, Japan % {jp.RYUKOKU} {246}
\\
$^{247}$Indian Institute of Science Education and Research, Kolkata, Mohanpur, West Bengal 741252, India % {IISER-KOL} {247}
\\
$^{248}$Universit\'e de Paris, 75006 Paris, France % {UNIPARIS} {248}
\\
$^{249}$Department of Physics, University of Notre Dame, Notre Dame, IN 46556, USA % {us.UNOTREDAME} {249}
\\
$^{250}$Centre national de la recherche scientifique, 75016 Paris, France % {CNRSPARIS} {250}
\\
$^{251}$Laboratoire Univers et Th\'eories, Observatoire de Paris, 92190 Meudon, France % {LUTMEUDON} {251}
\\
$^{252}$Observatoire de Paris, 75014 Paris, France % {OAPARIS} {252}
\\
$^{253}$Universit\'e PSL, 75006 Paris, France % {UNIPSL} {253}
\\
$^{254}$Institute of Physics of the Czech Academy of Sciences, 182 00 Praha 8, Czechia % {IPHYPRAHA} {254}
\\
$^{255}$Graduate School of Science and Technology, Niigata University, Nishi-ku, Niigata City, Niigata 950-2181, Japan % {jp.NIIGATA.PHYS} {255}
\\
$^{256}$Cornell University, Ithaca, NY 14850, USA % {Cornell} {256}
\\
$^{257}$Consiglio Nazionale delle Ricerche - Istituto dei Sistemi Complessi, I-00185 Roma, Italy % {CNRRM} {257}
\\
$^{258}$Korea Astronomy and Space Science Institute (KASI), Yuseong-gu, Daejeon 34055, Republic of Korea % {kr.KASI} {258}
\\
$^{259}$Hobart and William Smith Colleges, Geneva, NY 14456, USA % {HobartWilliamSmith} {259}
\\
$^{260}$International Institute of Physics, Universidade Federal do Rio Grande do Norte, Natal RN 59078-970, Brazil % {IIP-UFRN} {260}
\\
$^{261}$Museo Storico della Fisica e Centro Studi e Ricerche ``Enrico Fermi'', I-00184 Roma, Italy % {FERMI} {261}
\\
$^{262}$Dipartimento di Matematica e Fisica, Universit\`a degli Studi Roma Tre, I-00146 Roma, Italy % {DIMARM3} {262}
\\
$^{263}$Universit\`a di Trento, Dipartimento di Matematica, I-38123 Povo, Trento, Italy % {DIMATN} {263}
\\
$^{264}$University of California, Riverside, Riverside, CA 92521, USA % {UCR} {264}
\\
$^{265}$University of Washington, Seattle, WA 98195, USA % {UWashGravity} {265}
\\
$^{266}$Department of Electronic Control Engineering, National Institute of Technology, Nagaoka College, Nagaoka City, Niigata 940-8532, Japan % {jp.NAGAOKA.NIT} {266}
\\
$^{267}$INAF, Osservatorio Astronomico di Brera sede di Merate, I-23807 Merate, Lecco, Italy % {INAFME} {267}
\\
$^{268}$Departamento de Matem\'atica da Universidade de Aveiro and Centre for Research and Development in Mathematics and Applications, 3810-183 Aveiro, Portugal % {DEMAVEIR} {268}
\\
$^{269}$Marquette University, Milwaukee, WI 53233, USA % {Marquette} {269}
\\
$^{270}$Faculty of Science, Toho University, Funabashi City, Chiba 274-8510, Japan % {jp.TOHO} {270}
\\
$^{271}$Indian Institute of Technology, Palaj, Gandhinagar, Gujarat 382355, India % {IITGN} {271}
\\
$^{272}$Graduate School of Science and Technology, Gunma University, Maebashi, Gunma 371-8510, Japan % {jp.GunmaU} {272}
\\
$^{273}$Institute for Quantum Studies, Chapman University, Orange, CA 92866, USA % {us.ChapmanU} {273}
\\
$^{274}$Accelerator Laboratory, High Energy Accelerator Research Organization (KEK), Tsukuba City, Ibaraki 305-0801, Japan % {jp.KEK} {274}
\\
$^{275}$Faculty of Information Science and Technology, Osaka Institute of Technology, Hirakata City, Osaka 573-0196, Japan % {jp.OIT} {275}
\\
$^{276}$INAF, Osservatorio Astrofisico di Arcetri, I-50125 Firenze, Italy % {INAFFI} {276}
\\
$^{277}$Indian Institute of Technology Hyderabad, Sangareddy, Khandi, Telangana 502285, India % {IIT-Hydera} {277}
\\
$^{278}$Indian Institute of Science Education and Research, Pune, Maharashtra 411008, India % {IISER-Pune} {278}
\\
$^{279}$Istituto di Astrofisica e Planetologia Spaziali di Roma, 00133 Roma, Italy % {IAPSRM} {279}
\\
$^{280}$Department of Space and Astronautical Science, The Graduate University for Advanced Studies (SOKENDAI), Sagamihara City, Kanagawa 252-5210, Japan % {jp.SOKEN.JAXA} {280}
\\
$^{281}$Andrews University, Berrien Springs, MI 49104, USA % {Andrews} {281}
\\
$^{282}$Research Center for Space Science, Advanced Research Laboratories, Tokyo City University, Setagaya, Tokyo 158-0082, Japan % {jp.TCU} {282}
\\
$^{283}$Institute for Cosmic Ray Research (ICRR), Research Center for Cosmic Neutrinos (RCCN), The University of Tokyo, Kashiwa City, Chiba 277-8582, Japan % {jp.ICRR.CCN} {283}
\\
$^{284}$Department of Physics, Kyoto University, Sakyou-ku, Kyoto City, Kyoto 606-8502, Japan % {jp.KYOTO.PHYS} {284}
\\
$^{285}$Yukawa Institute for Theoretical Physics (YITP), Kyoto University, Sakyou-ku, Kyoto City, Kyoto 606-8502, Japan % {jp.YITP} {285}
\\
$^{286}$Dipartimento di Scienze Aziendali - Management and Innovation Systems (DISA-MIS), Universit\`a di Salerno, I-84084 Fisciano, Salerno, Italy % {DISAMISSA} {286}
\\
$^{287}$Van Swinderen Institute for Particle Physics and Gravity, University of Groningen, 9747 AG Groningen, Netherlands % {UNIGRON} {287}
\\
$^{288}$Faculty of Science, Department of Physics, The Chinese University of Hong Kong, Shatin, N.T., Hong Kong % {hk.CUHK} {288}
\\
$^{289}$Vrije Universiteit Brussel, 1050 Brussel, Belgium % {VRIJEBRUS} {289}
\\
$^{290}$Applied Research Laboratory, High Energy Accelerator Research Organization (KEK), Tsukuba City, Ibaraki 305-0801, Japan % {jp.KEK.ARL} {290}
\\
$^{291}$Department of Communications Engineering, National Defense Academy of Japan, Yokosuka City, Kanagawa 239-8686, Japan % {jp.NDAJ} {291}
\\
$^{292}$Department of Physics, University of Florida, Gainesville, FL 32611, USA % {us.UFLORIDA} {292}
\\
$^{293}$Department of Information and Management Systems Engineering, Nagaoka University of Technology, Nagaoka City, Niigata 940-2188, Japan % {jp.NAGAOKATECH} {293}
\\
$^{294}$Tata Institute of Fundamental Research, Mumbai 400005, India % {TIFR} {294}
\\
$^{295}$Eindhoven University of Technology, 5600 MB Eindhoven, Netherlands % {UNIEINDH} {295}
\\
$^{296}$Department of Physics and Astronomy, Sejong University, Gwangjin-gu, Seoul 143-747, Republic of Korea % {kr.SEJONG} {296}
\\
$^{297}$Concordia University Wisconsin, Mequon, WI 53097, USA % {CUW} {297}
\\
$^{298}$Department of Electrophysics, National Yang Ming Chiao Tung University, Hsinchu, Taiwan % {tw.NCTU} {298}
\\
$^{299}$Department of Physics, Rikkyo University, Toshima-ku, Tokyo 171-8501, Japan % {jp.RIKKYO} {299}
\\
}

\pubyear{2022}

\begin{document}
\begin{titlepage}
\maketitle
\end{titlepage}

\newpage
\label{firstpage}
\pagerange{\pageref{firstpage}--\pageref{lastpage}}
\begin{abstract}

We describe a search for gravitational waves from compact binaries with at least one component with mass $0.2 \thinspace M_\odot$--$1.0 \thinspace M_\odot$ and mass ratio $q\geq 0.1$ in Advanced LIGO and Advanced Virgo data collected between 1 November 2019, 15:00 UTC and 27 March 2020, 17:00 UTC.  No signals were detected. The most significant candidate has a false alarm rate of \FARa{}. We estimate the sensitivity of our search over the entirety of Advanced LIGO's and Advanced Virgo's third observing run, and present the most stringent limits to date on the merger rate of binary black holes with at least one  subsolar-mass component. We use the upper limits to constrain two fiducial scenarios that could produce subsolar-mass black holes: primordial black holes (PBH) and a model of dissipative dark matter. The PBH model uses recent prescriptions for the merger rate of PBH binaries that include a rate suppression factor to effectively account for PBH early binary disruptions.
If the PBHs are monochromatically distributed, we can exclude a dark matter fraction in PBHs $f_\mathrm{PBH} \gtrsim \, \fPBHlim $ (at 90\% confidence) in the probed subsolar-mass range. However, if we allow for broad PBH mass distributions we are unable to rule out $f_{\rm PBH} = 1$. For the dissipative model, where  the dark matter has chemistry that allows a small fraction to cool and collapse into black holes, we find an upper bound $f_{\rm DBH} < 10^{-5}$ on the fraction of atomic dark matter collapsed into black holes.\\

\end{abstract}

%\maketitle

% Select between one and six entries from the list of approved keywords.
% Don't make up new ones.

\section{Introduction}\label{sec:intro}

The Advanced LIGO~\citep{TheLIGOScientific:2014jea} and Advanced Virgo~\citep{TheVirgo:2014hva} detectors have completed three observing runs, O1, O2, and O3 (split into O3a and O3b), since the first observation of gravitational waves from a binary black hole (BBH) coalescence~\citep{Abbott:2016blz}. The collected data have been analyzed by the LIGO--Virgo--KAGRA (LVK) Collaboration~\citep{LVKarticle} in successive versions of the Gravitational Wave Transient Catalog~\citep[GWTC;][]{TheLIGOScientific:2016pea,LIGOScientific:2018mvr,LIGOScientific:2020ibl,LIGOScientific:2021usb,LIGOScientific:2021djp}, which report a total of 90 candidate gravitational-wave (GW) events from the coalescence of compact binary systems with a probability of astrophysical origin $>$ 0.5. 
Several additional candidates of compact binary signals have also been included in independent catalogs~\citep{Nitz:2018imz,Magee:2019vmb,Venumadhav:2019tad,Venumadhav:2019lyq,Nitz:2019hdf,Nitz:2021uxj,Nitz:2021zwj,Olsen:2022pin} after analyzing the publicly released strain data~\citep{RICHABBOTT2021100658}.    
These detections have revealed features in the population of coalescing objects that revolutionize our previous understanding of astrophysics and stellar evolution~\citep{Mandel:2018hfr,Spera:2022byb}. 
The masses of many black holes (BHs) detected in GWs are much larger than 
 those of the BHs observed in X--ray binaries~\citep{Bailyn:1997xt,Ozel:2010su,Farr:2010tu,Fishbach:2021xqi} and some signals, such as GW190521~\citep{Abbott:2020tfl,Abbott:2020mjq}, have primary component masses within the predicted pair-instability mass gap~\citep{Woosley:2016hmi,Farmer:2019jed}. On the other side of the mass range are events like GW190425~\citep{Abbott:2020uma}, whose total mass is substantially larger than any known Galactic neutron star binary \citep{Farrow:2019xnc,LIGOScientific:2019eut}, and events like GW190814~\citep{Abbott:2020khf,LIGOScientific:2020kqk}
 and GW200210${}_{-}$092254~\citep{LIGOScientific:2021djp} that are also atypical due to their highly asymmetric masses
 and the properties of their light components~\citep{Zevin:2020gma}. While open questions remain, GWs have provided a unique census of the population of black holes in binaries in our Universe~\citep{LIGOScientific:2021psn}.

Current models of stellar evolution predict that white dwarfs that end their thermonuclear burning with a mass greater than the Chandrasekhar limit~\citep{1931ApJ.74.81C,10.1093/mnras/95.3.207,Suwa:2018uni,Muller:2018utr,Ertl:2019zks}  will  collapse to form either a neutron star or a supersolar-mass black hole.
Since there are no standard astrophysical channels that produce subsolar-mass objects more compact than white dwarfs,
the detection of a subsolar-mass (\SSM{}) compact object would indicate the presence of a new formation mechanism alternative to usual stellar evolution. 

Given the still-unknown nature of 84\% of the matter in the Universe~\citep{Planck:2018vyg}, it is reasonable to consider whether the DM might be composed of, or produce, distinct populations of compact objects.
Primordial black holes (PBHs), postulated to form from the collapse of large overdensities in the early Universe~\citep{Zeldovich:1967lct,Hawking:1971ei,Carr:1974nx,Chapline:1975ojl}, are  candidates to form  at least a fraction of the dark matter (\DM{})  while providing an explanation to several open problems in astrophysics and cosmology~\citep{Barrow:1990he,Bean:2002kx,Kashlinsky:2016sdv,Clesse:2017bsw}. Soon after the first BBH coalescence was observed, it was suggested~\citep{Bird:2016dcv,Clesse:2016vqa,Sasaki:2016jop} that the detected BHs could have a primordial origin. 
Large primordial fluctuations at small scales generated during inflation can produce PBHs~\citep{Carr:1993aq,Ivanov:1994pa,Kim:1996hr,GarciaBellido:1996qt}, though other processes in the early Universe, like bubble nucleation and domain walls~\citep{Garriga:2015fdk}, cosmic string loops, and scalar field instabilities~\citep{Khlopov:1985jw,Cotner:2017tir} can also be sources of overdensities that eventually collapse to produce PBHs~\citep{Khlopov:2008qy,Carr:2020gox,Carr:2020xqk,Villanueva-Domingo:2021spv}. 
The thermal history of the Universe can further enhance the formation of PBH at different scales~\citep{Carr:2019kxo}.
For example, the quark--hadron (QCD) transition significantly reduces the radiation pressure of the plasma, so that a uniform primordial enhancement stretching across the QCD scale will generate a distribution of PBH masses that is sharply peaked around a solar mass~\citep{Byrnes:2018clq} as well as a broader mass distribution at both larger and smaller masses that could explain some of the GW observations~\citep{Clesse:2020ghq,Jedamzik:2020omx,Jedamzik:2020ypm,Chen:2021nxo,Juan:2022mir,Franciolini:2022pav}. In particular, GW events in the \SSM{} range could be used to probe mergers involving PBH black holes from a QCD enhanced peak.

Models of particle dark matter can also produce compact objects either from an interaction of dark matter with Standard Model particles,  such as boson stars or neutron stars transmuted into black holes due to DM accretion~\citep{Dasgupta:2020mqg,Kouvaris:2018wnh,Kouvaris:2010jy,deLavallaz:2010wp,PhysRevD.40.3221,Bramante:2015dfa,Bramante:2014zca,Bramante:2017ulk,Takhistov:2017bpt,Takhistov:2020vxs}, or directly from the gravitational collapse of dissipative DM~\citep{Ryan:2021dis, Chang:2018bgx, Shandera:2018xkn, Choquette:2018lvq, Latif:2018kqv, DAmico:2017lqj, Essig:2018pzq, Hippert2021}. \DM{} black holes (DBHs) may form in the late universe if DM has a sufficiently rich particle content to allow dissipation and collapse of \DM{} into compact structures.  While these
mechanisms generically produce black holes that overlap the standard astrophysical population, under specific assumptions they may also be able to create SSM compact objects. 

Searches for compact binaries with at least one component below $1\thinspace M_\odot$ have been carried out using both Initial LIGO~\citep{Abbott:2005pf,Abbott:2007xi}, and Advanced LIGO and Advanced Virgo data~\citep{LIGOScientific:2018glc,LIGOScientific:2019kan,LIGOScientific:2021job,Nitz:2021vqh,Nitz:2022ltl,Nitz:2020bdb,Phukon:2021cus,Nitz:2021mzz}. No firm detections were reported in any of these analyses. We describe and present the results of the search for the GWs from binary systems with at least one \SSM{} component down to $0.2 \msun$, using data from the second part of the third observing run (O3b) in Sec.~\ref{sec:search}. We find no unambiguous GW candidates. The null result, combined with our previous analysis of the first part of the third observing run~\citep[O3a;][]{LIGOScientific:2021job}, allows us to set in Sec.~\ref{sec:injections} upper limits on the merger rate
of binaries with one SSM component, as function of the chirp mass and in the $m_1$--$m_2$ plane.

{These new upper limits on the merger rate can be used to constrain any model that might generate compact objects in the SSM range. As illustrative examples, we derive in Sec.~\ref{sec:constraints} new constraints on two particular scenarios, PBHs and a model of DBHs.}
For PBH models, we calculate the merger rate of SSM binaries taking into account the early~\citep{Hutsi:2020sol} and late binary formation scenarios~\citep{Clesse:2020ghq,Phukon:2021cus}, and we reevaluate the constraints on PBH DM models with monochromatic (delta-function) and extended
mass distributions.
We update the PBH merger rate model of previous LVK works~\citep{Abbott:2018oah,Authors:2019qbw,LIGOScientific:2021job} with additional physics to allow for binary disruption and find that the constraints on monochromatically distributed PBHs are weakened.
We also {consider} broad PBH mass functions such as those of thermal history scenarios of PBHs and find that they are not significantly constrained in the SSM range by the present LVK data.
For DBHs, we constrain a simple atomic dark matter model where DM consists of two oppositely charged dark fermions interacting via a dark photon~\citep{Shandera:2018xkn}. This model has been estimated to produce a sizeable population of SSM black holes if the heavier of the fermions, $X$, is more massive than the Standard Model proton~\citep{Shandera:2018xkn}; the fermion mass range previously probed was
$0.66 \thinspace \mathrm{GeV}/c^2< m_X < 8.8 \thinspace \mathrm{GeV}/c^2$~\citep{LIGOScientific:2021job,Singh:2020wiq}.  We obtain improved constraints on the fraction of DM in DBHs as a function of the minimum mass of the DBHs. In Sec.~\ref{sec:conc} we summarize our findings and discuss prospects for Advanced LIGO and Advanced Virgo's fourth observing run.

\section{Search}
\label{sec:search}

\begin{table*}
\caption{The triggers with a FAR $<2$ yr$^{-1}$ in at least one search pipeline. We include the search-measured parameters associated with each candidate: $m_1$ and $m_2$, the redshifted component masses, and $\chi_{1}$ and $\chi_{2}$, the dimensionless component spin. The parameters shown in the table are the ones reported by the search where the trigger is identified with the lowest FAR. H, L, and V denote the Hanford, Livingston, and Virgo interferometers, respectively. The dashes in the ``V SNR'' column mean that no single-detector trigger was found in Advanced Virgo. The network SNR is computed by adding the SNR of single detector triggers in quadrature.}
\begin{tabular}{c c c c c c c c c c c}
FAR [yr$^{-1}$] & Pipeline &  GPS time & $m_1$ $[M_{\odot}]$ & $m_2$ $[M_{\odot}]$ & $\chi_{1}$ & $\chi_{2}$ & H SNR & L SNR & V SNR & Network SNR \\
\hline
0.20 & \texttt{GstLAL} & 1267725971.02 & 0.78 & 0.23 & 0.57 & 0.02 & 6.31 & 6.28 &  -  & 8.90\\
1.37 & \texttt{MBTA}   & 1259157749.53 & 0.40 & 0.24 & 0.10 &$-0.05$ & 6.57 & 5.31 & 5.81 & 10.25\\
1.56 & \texttt{GstLAL} & 1264750045.02 & 1.52 & 0.37 & 0.49 & 0.10 & 6.74 & 6.10 &  -  & 9.10\\
\end{tabular}
\label{tab:candidate}
\end{table*}

The \SSM{} search analyzes data collected during O3b, covering the period from 1 November 2019 1500 UTC to 27 March 2020 1700 UTC. The characterization and calibration of data and the non-linear removal of spectral lines follow the same methods as in our O3a analyses~\citep{LIGOScientific:2021usb,LIGOScientific:2020ibl,LIGOScientific:2021job}. 

The analysis is performed by using three matched-filtering pipelines: \texttt{GstLAL}~\citep{Messick:2016aqy,Sachdev:2019vvd,Hanna:2019ezx}, \texttt{MBTA}~\citep{Aubin:2020goo} and \texttt{PyCBC}~\citep{Allen:2005fk,Allen:2004gu,Canton:2014ena,Usman:2015kfa,Nitz:2017svb,Davies:2020tsx}. These analyses correlate the data with a bank of templates that model the gravitational-wave signals expected from binaries in quasi-circular orbit. All search pipelines use the same template banks and the same setup as for the O3a \SSM{} analysis~\citep{LIGOScientific:2021job}. Templates are generated using the \texttt{TaylorF2} waveform~\citep{Sathyaprakash:1991mt, Blanchet:1995ez, Poisson:1997ha, Damour:2001bu, Mikoczi:2005dn, Blanchet:2005tk, Arun:2008kb,Buonanno:2009zt,Bohe:2013cla,Bohe:2015ana,Mishra:2016whh} and include phase terms {up} to 3.5 post-Newtonian order, but no amplitude corrections.  We estimate the GW emission starting at a frequency of $45$ Hz to limit the computational cost of the search; we estimate that this reduces the network average signal-to-noise ratio (SNR) by \SNRLossFromFlow. The template bank was constructed using a geometric placement algorithm~\citep{Harry:2013tca}. 
The bank is designed to recover binaries with (redshifted) primary mass $m_1 \in$ [\MinPrimaryMass, \MaxPrimaryMass] \msun and  secondary mass $m_2 \in$ [\MinSecondaryMass, \MaxSecondaryMass] \msun . The lower mass bound is set for consistency with previous searches~\citep{Abbott:2018oah,Authors:2019qbw,LIGOScientific:2021job} and to limit the computational cost of the search. We additionally limit the binary mass ratio, $q \equiv m_2 / m_1$, with $m_2 \leq m_1$, to range from $\MinMassRatio < q < 1.0$. We include the effect of spins aligned with the orbital angular momentum. For masses of a binary component larger than $0.5 \thinspace M_\odot$ we allow for a dimensionless component spin ($\chi_{1,2} = |\textbf{S}_{1,2}|/m_{1,2}^2$, with $\textbf{S}_{1,2}$ the angular momentum of the compact objects) up to \MaxHighMassComponentSpin{}, while for compact objects with masses less than or equal to $0.5 \thinspace M_\odot$, we limit the maximum dimensionless
spin to \MaxLowMassComponentSpin{}. The restriction on component spins is chosen to reduce the computational cost of the analyses~\citep{LIGOScientific:2021job}. We set a minimum match~\citep{Owen:1995tm} of \MinMatch{} to ensure that no more than {10\%} of astrophysical signals can be missed due to the discrete sampling of the parameter space.

We report in Table~\ref{tab:candidate} the most significant candidates down to the threshold false alarm rate (FAR) of FAR $<2 \thinspace \mathrm{yr}^{-1}$. We do not apply a trials factor to our analysis. We identify only three triggers that pass this threshold in at least one pipeline. Visual inspection of the data around the time of the triggers indicate no data quality issues that would point to a definitive instrumental origin of the candidates. However, the number of triggers with their estimated FAR is consistent with what we would expect if no astrophysical signal was present in the data, given that the duration of O3b is $0.34$ yr and that three pipelines are being used. The most significant candidate has a FAR of \FARa{}, which assuming a Poisson distribution for the background triggers and an observing time of $0.34$ yr, corresponds to a p-value of 6.6\%. We conclude that there is no statistically significant evidence for the detection of a GW from a \SSM{} source.

\section{Sensitivity and rate limits}

\label{sec:injections}
The absence of significant candidates in O3b allows us to characterize the sensitivity of our search and to set upper limits on the merger rate of such binary systems. We estimate the sensitive volume--time $\langle VT \rangle$ over all of O3.
We find the sensitivity of each of the three pipelines introduced in Sec.~\ref{sec:search} with a common set of simulated signals in real data, generated using the precessing post-Newtonian waveform model \texttt{SpinTaylorT5} \citep{PhysRevD.84.084037}, with source component masses sampled from log-uniform distributions with primary masses in range $(0.19, 11.0) ~\msun$ and secondary masses in range $(0.19, 1.1) ~\msun$. The injection's component spins are distributed isotropically with dimensionless spin magnitudes going up to $0.1$. The injections are distributed uniformly in comoving volume up to a maximum redshift of $z = 0.2$, at which the sensitivity of the search has been checked to be negligible.
We injected a total of approximately $2$ million simulated signals, spaced $15$ s apart, spanning all O3.

The sensitivity of each search pipeline is estimated by computing the sensitive volume--time of the search:
\begin{equation} 
\langle VT \rangle = \epsilon ~V_{\mathrm{inj}} ~T \,, 
\label{eqVT}
\end{equation}
where $\epsilon$ is the efficiency, defined as the ratio of recovered to total injections in the data in the source frame mass bin of interest, $T$ is the analyzed time, and $V_{\text{inj}}$ is the comoving volume at the farthest injected simulation.
Each pipeline uses all injections with $q > 0.05$. We evaluate the uncertainties at $90\%$ confidence interval on the sensitive volume--time estimate~\citep{Tiwari:2017ndi} and consider binomial errors on the efficiency $\epsilon$, given by
\begin{equation}
    \delta \left( VT \right) = 1.645 ~\sqrt{\frac{\epsilon \left(1 - \epsilon \right)}{N_{\text{\rm{inj}}}}} ~V_{\rm{inj}} ~T \,,
\label{eq:eqVTError}
\end{equation}
where $N_{\text{inj}}$ are the total injections in the considered mass range.

\begin{figure}
\includegraphics[width=\columnwidth]{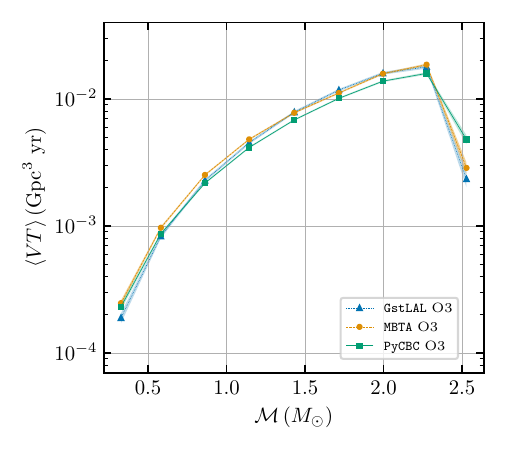}
\caption{\label{fig:VTO3} Sensitive volume--time as a function of the source frame chirp mass in data from O3, obtained through the analysis of the set of common injections (blue triangles with dotted lines, orange circles with dashed lines, and green squares with continuous lines). The statistical errors are evaluated at $90\%$ confidence interval, following Eq.~(\ref{eq:eqVTError}) and represented by the shaded areas.
}
\end{figure}

We use the FAR of the most significant candidate in O3 for each pipeline to estimate the upper limit on the merger rate in accordance with the loudest event statistic formalism~\citep{Biswas:2007ni}. The FAR thresholds used were \GstLALbestFAR, \MBTAbestFAR and \PyCBCbestFAR~\citep{LIGOScientific:2021job} for \texttt{GstLAL}, \texttt{MBTA} and \texttt{PyCBC}, respectively. By omitting a trials factor in our analysis, we obtain a conservative upper limit on the sensitive $\langle VT \rangle$ of the searches. Though \texttt{MBTA} and \texttt{PyCBC} results use the full injection set, \texttt{GstLAL} analyzed a subset; the uncertainties in $\langle VT \rangle$ shown in Fig.~\ref{fig:VTO3} are therefore larger for \texttt{GstLAL}. 

To lowest order, the inspiral of a binary depends sensitively on the chirp mass of the system~\citep{Blanchet:2013haa}, which is defined as $\mathcal{M} \equiv (m_1 m_2)^{3/5} / (m_1+m_2)^{1/5}$. Therefore, we split the population into nine equally spaced chirp mass bins in the range $0.16 M_\odot \leq \mathcal{M} \leq 2.72 M_\odot$ to determine the $\langle VT \rangle$ as a function of the chirp mass, shown in Fig.~\ref{fig:VTO3}. The highest chirp mass bin of this search exhibits a drop in sensitivity as the component masses contained within this bin are beyond the redshifted component masses covered by the template bank (Sec.~\ref{sec:search}). As a consequence, there is a drop in efficiency and smaller $\langle VT \rangle$ values in that region.
The sensitivity estimates obtained from the analysis of O3a data with the common injection set are consistent with the ones reported in our previous work~\citep{LIGOScientific:2021job}.

The null result from O3 yields $\langle VT \rangle$ values approximately $2$ times larger than those obtained for O3a, in agreement with the expected increase in observing time.  The sensitive hypervolumes of the searches presented in GWTC-3~\citep{LIGOScientific:2021djp} for chirp masses of 1.3 $M_\odot$ and 2.3 $M_\odot$ are comparable to those in~Fig.~\ref{fig:VTO3} even though the mass ratio bounds of the two populations are different.

\begin{figure}
\includegraphics[width=\columnwidth]{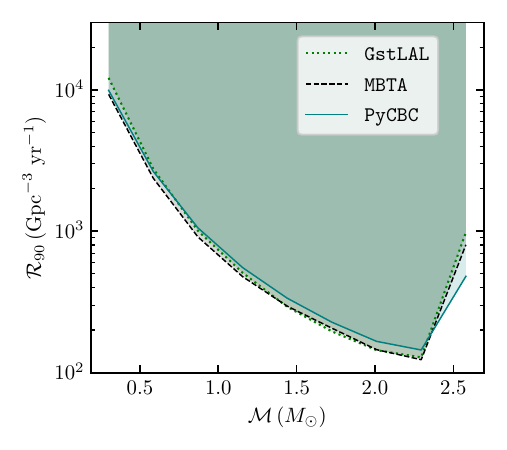}
\caption{\label{fig:rateO3mc} Merger rate limits as function of the source frame chirp mass of the binary system, in data from the full O3. The dotted, dashed and solid lines represent the $90\%$ confidence limits obtained by \texttt{GstLAL}, \texttt{MBTA} and \texttt{PyCBC}, respectively.}
\end{figure}

\begin{figure}
\includegraphics[width=\columnwidth]{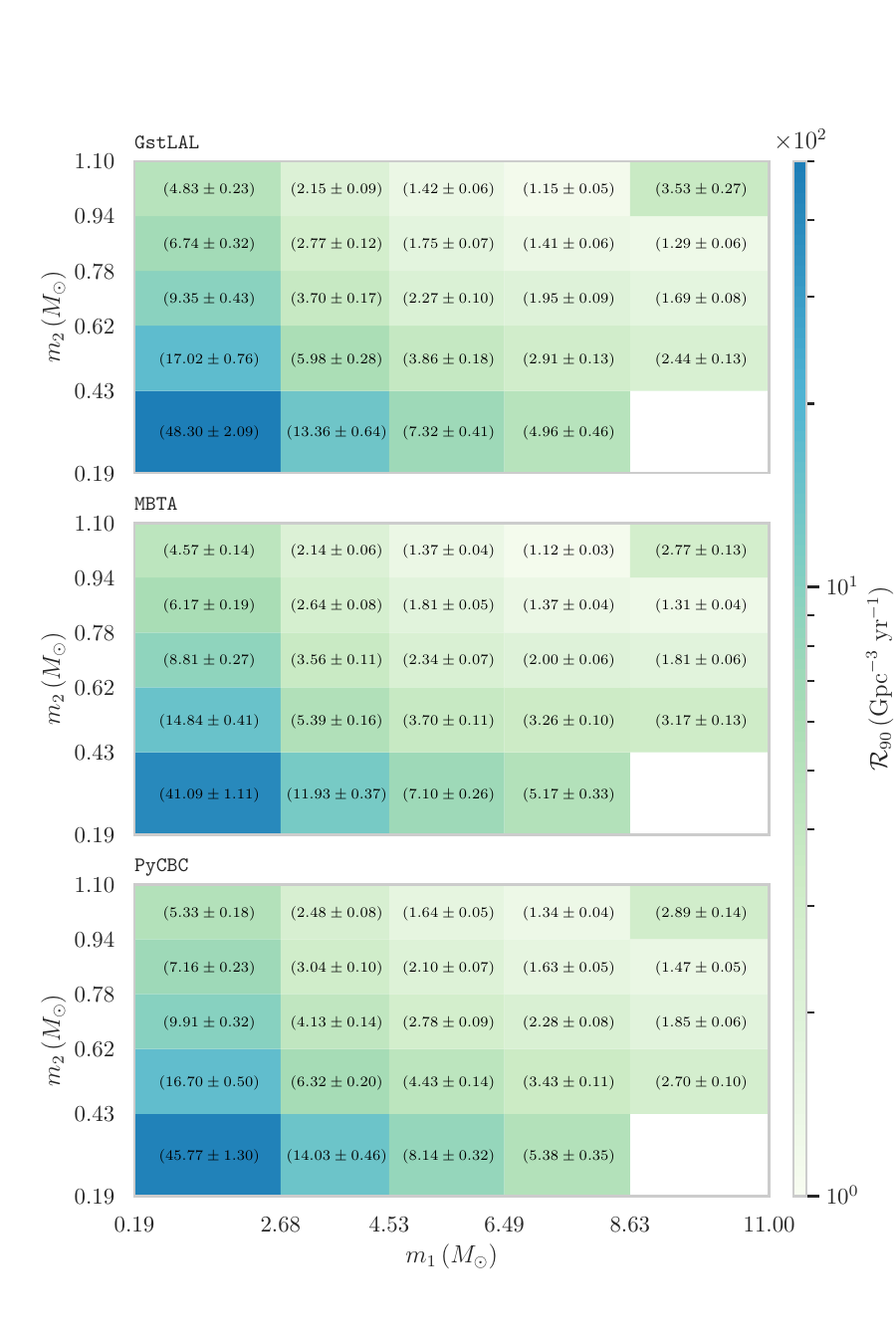}
\caption{\label{fig:rateO3m1m2} Merger rate limits in the source frame $m_1$--$m_2$ plane, in data from the full O3 for the three pipelines. The error bars in each panel are given at the $90\%$ confidence interval, following Eq.~\ref{eq:eqVTError}.}
\end{figure}

Given the obtained sensitive volume and the absence of significant detection, one can infer merger rate limits. Treating each bin, $i$, as a different population, we computed an upper limit on the binary merger rate to $90\%$ confidence~\citep{Biswas:2007ni}:
\begin{equation} 
\mathcal{R}_{90,i} = \frac{2.3}{\langle VT \rangle_i}. \label{eqR90}
\end{equation}
We show in Fig.~\ref{fig:rateO3mc} and in Fig.~\ref{fig:rateO3m1m2} the upper limits on the binary merger rate as function of the chirp mass and in the source $m_1$\textendash $m_2$ plane, respectively.

\section{Constraints on dark matter models}
\label{sec:constraints}

The upper limits that we infer from our null result can generically be used to constrain models that predict an observable population of binaries with at least one \SSM{} component. We connect our results to two possible sources of \SSM{} black holes: PBHs and DBHs. We parameterize our constraints in terms of the fraction of the dark matter that can be comprised of compact objects under each model.

\subsection{Primordial Black Holes}

The abundance and mass distribution of PBHs depend on the details of their particular formation mechanism. The primordial power spectrum generated during inflation must have sufficiently large fluctuations on small scales for PBHs formation, while keeping the fluctuations small at the scale of the observed cosmic microwave background anisotropies~\citep{Cole:2022xqc}. This is possible in several two-field models of inflation~\citep{Clesse:2015wea,Braglia:2020eai,Zhou:2020kkf,DeLuca:2020agl}, single-field models with a non slow-roll regime due to specific features in the inflation dynamics ~\citep{Garcia-Bellido:2017mdw,Ezquiaga:2017fvi}, and by the enhancement of fluctuations at small scales due to quantum diffusion ~\citep{Pattison:2017mbe,Ezquiaga:2019ftu}, which provide recent examples of inflationary scenarios that can produce PBHs in the SSM range.

The probability of matter fluctuations to collapse into PBHs is enhanced by the decrease of the radiation pressure as different particles become non-relativistic along the thermal history of the Universe~\citep{Carr:2019kxo}. In particular, a peak around a solar mass is expected due to the QCD transition, although its exact position and height depend on the characteristics of the matter fluctuations at those scales~\citep{Byrnes:2018clq}. Furthermore, the probability of binary formation and thus estimates of the event rates depends on the clustering of PBHs and the cluster dynamics. This remains an area of active study~\citep{Raidal:2018bbj,Trashorras:2020mwn,Jedamzik:2020ypm}.
All these uncertainties make our predictions on the DM fraction of PBHs very sensitive to the particular choice of the model parameters~\citep{Escriva:2022bwe,Franciolini:2022tfm}.

We update the theoretical merger rate of PBHs used in previous LVK searches~\citep{Abbott:2018oah,Authors:2019qbw,LIGOScientific:2021job}. We approximate the merger rates of early PBH binaries (EBs) formed in the radiation-dominated era with the approximations provided by~\citet{Hutsi:2020sol,Chen:2018czv,Ali-Haimoud:2017rtz} and numerically validated with N-body simulations in~\citet{Raidal:2018bbj},
\begin{align}
\frac{{\rm d}\mathcal R^{\rm PBH}} {{\rm d} \ln m_1 {\rm d} \ln m_2} &= 1.6 \times 10^6\thinspace \mathrm{Gpc}^{-3} \thinspace \mathrm{yr}^{-1} \times f_{\rm sup} f_{\rm PBH}^{53/37} \nonumber 
  f(m_1) \times \\ &f(m_2) \left( \frac{m_1+m_2}{M_\odot} \right)^{-32/37} \left[ \frac{m_1 m_2}{(m_1+m_2)^2}\right]^{-34/37},  \label{eq:REB}
 \end{align}
where $f_{\rm PBH}$ denotes the \DM{} density fraction made of PBHs and $f(m)$ is the normalized PBH density distribution.  We neglect the redshift dependence in the merger rates, since the current generation of ground-based interferometers is only sensitive to BBHs with at least one SSM component at low redshifts. The main difference, compared to the theoretical rates predicted by~\citet{Sasaki:2016jop} that were used in previous LVK searches, comes from a rate suppression factor $f_{\rm sup}$ that effectively accounts for PBH binary disruptions by early forming clusters due to Poisson fluctuations in the initial PBH separation, by matter inhomogeneities, and by nearby PBHs~\citep{Suyama:2019cst,Matsubara:2019qzv}.  For instance, if PBHs have all the same mass or  a strongly peaked mass function and significantly contribute to the dark matter, one gets $f_{\rm sup} \approx 2.3 \times 10^{-3} f_{\rm PBH}^{-0.65}$, so the merger rates are highly suppressed~\citep{Hutsi:2020sol}.  As a result, the limits on $f_{\rm PBH}$ are much less stringent than previously estimated.  Data from O2 still allow for $f_{\rm PBH} = 1$ in a scenario where all the PBHs have the same mass.  Though monochromatically distributed PBHs are unrealistic, they provide a useful approximation for models with a highly peaked distribution, e.g., as predicted from PBH scenarios with sharp QCD transitions~\citep{Carr:2019kxo}. Given the still large uncertainties and possible caveats for the merger rate prescriptions of early binaries, we also considered {the case where merger rates entirely come from} late PBH binaries (LBs) formed dynamically inside PBH clusters seeded by the above-mentioned Poisson fluctuations that grow in the matter-dominated era and lead to the formation of PBH clusters, following~\citet{Clesse:2020ghq,Phukon:2021cus}.  {This allows us to illustrate the important variations in the PBH limits obtained for different binary formation scenarios. } 

\begin{figure}
\hspace*{-1mm}
\includegraphics[width=\columnwidth]{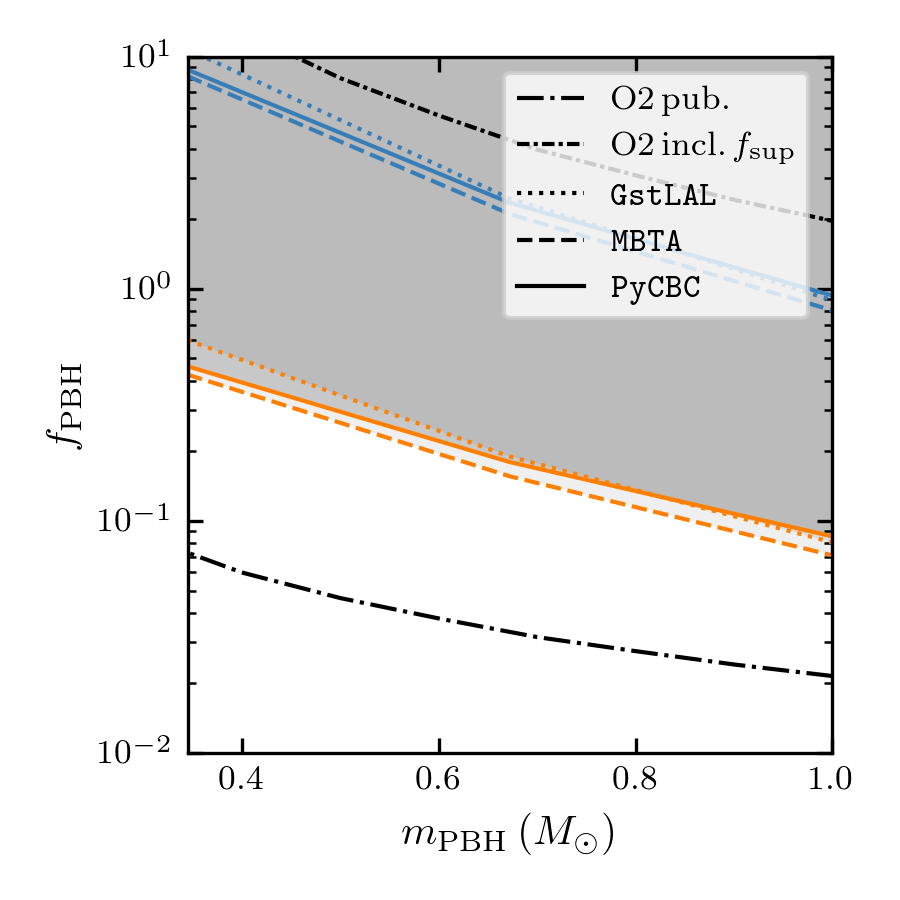}
\caption{\label{fig:fPBH} Constraints on \DM{} fraction of PBHs, $f_\mathrm{PBH}$, for a monochromatic mass function and assuming the merger rates for early PBH binaries from~\citet{Hutsi:2020sol} (orange) and late PBH binaries from~\citet{Phukon:2021cus} (blue). Shown in black are results for \SSM{} searches in O2~\citep{LIGOScientific:2019kan} with and without the rate suppression factor $f_{\rm sup}$. For the first time, $f_{\rm PBH} = 1$ for early binaries is excluded in the whole \SSM{} range probed by this search.  
}
\end{figure}

For a monochromatic PBH mass distribution, we derive new limits on $f_{\rm PBH}$ in the \SSM{} range, shown in Fig.~\ref{fig:fPBH}, for both EBs and LBs.  While the scenario of \DM{} entirely made of PBHs with the same mass was not totally excluded by previous searches, after O3 it becomes strongly disfavored up to $1M_\odot$, with $f_{\rm PBH} < 0.6$ around $0.3 M_\odot$ and $f_{\rm PBH} < 0.09 $ at $1 M_\odot$.  For LBs only, we do not find yet significant limits, since we do not restrict $f_{\rm PBH}$ to be lower than one.

For unequal mass BBH, the merger rates are more uncertain and model dependent, but one can obtain a limit on an effective parameter
\begin{equation}
    F_{\rm PBH} \equiv \left( \frac{f_{\rm sup}}{2.3 \times 10^{-3}} \right) f(m_1) f(m_2) f_{\rm PBH}^{53/37}~,
\end{equation}
in such a way that it corresponds to the product of $f(m_2)$ and $f(m_1)$ in a scenario where $f_{\rm PBH} \approx 1$. This allows us to establish model-independent limits on PBHs {since $F_{\rm PBH}$ encompasses all the uncertainties on the mass distribution and rate suppression}, by using the limits shown in Fig. \ref{fig:rateO3m1m2} and the rates of Eq.~(\ref{eq:REB}) but neglecting their variations in individual mass bins.  We find that the limits on $F_{\rm PBH}$ is sensitive to the location in the $m_1$--$m_2$ plane. These can be used to constrain $f_{\rm PBH}$ for arbitrary mass functions.  For models with $f_{\rm PBH} = 1$ and a peak above $1M_\odot$, these restrict the possible distribution of BHs in the SSM range. We find that {some} representative distributions with QCD-enhanced features~\citep{Byrnes:2018clq,Carr:2019kxo,DeLuca:2020agl,Jedamzik:2020omx} become  constrained in the range $f_{\rm PBH} \approx $ (0.1--1). \SSM{} searches are therefore complementary to searches in the solar mass range in order to distinguish PBH mass functions that are viable from those that are more constrained.

\subsection{Dark black holes}

If all or some of the \DM{} has rich enough particle content to dissipate kinetic energy and cool, then compact objects made from \DM{} may form through gravitational collapse of the dark gas~\citep{Shandera:2018xkn}. The particle content of the \DM{} allows \SSM{} black holes if, for example, there is a cosmologically dominant heavy fermion analogous to the proton but with mass greater than $938$ MeV/$c^2$. In that case, the Chandrasekhar limit for \DM{} black holes is lower than that for Standard Model matter. Constraints on \SSM{} black holes in mergers then constrain formation channels for \DM{} black holes in the detectable mass range, bounding the total cooling rate (total dissipation) of the dark sector~\citep{Singh:2020wiq}. 

Here we consider a population of DBHs formed within a particular dissipative scenario, the atomic \DM model~\citep{Ackerman2009,Kaplan2010,Feng2009}, with a power-law distribution of masses modeled after observations and simulations of Population III stars~\citep{2013MNRAS.433.1094S,Greif:2011it,Hartwig:2016nde}. We derive the posterior probability for the fraction of dissipative \DM{} that can be in black holes, the lower and upper limits of the DBH mass distribution, and the power-law slope, using the sensitive volume from the \SSM{} search and modelled rates for DBH mergers \citep{Shandera:2018xkn,Singh:2020wiq}. The posterior is marginalized over the parameters that characterize the distribution, including the power-law slope and the upper limit of the distribution to obtain the constraints on the fraction of dissipative \DM{}  that can be in black holes, $f_{\rm DBH}$, together with the lower limit of the DBH distribution $M_{\rm min}^{\rm DBH}$, as done in \cite{Singh:2020wiq,LIGOScientific:2021usb} previously.

\begin{figure}
\includegraphics[width=\columnwidth]{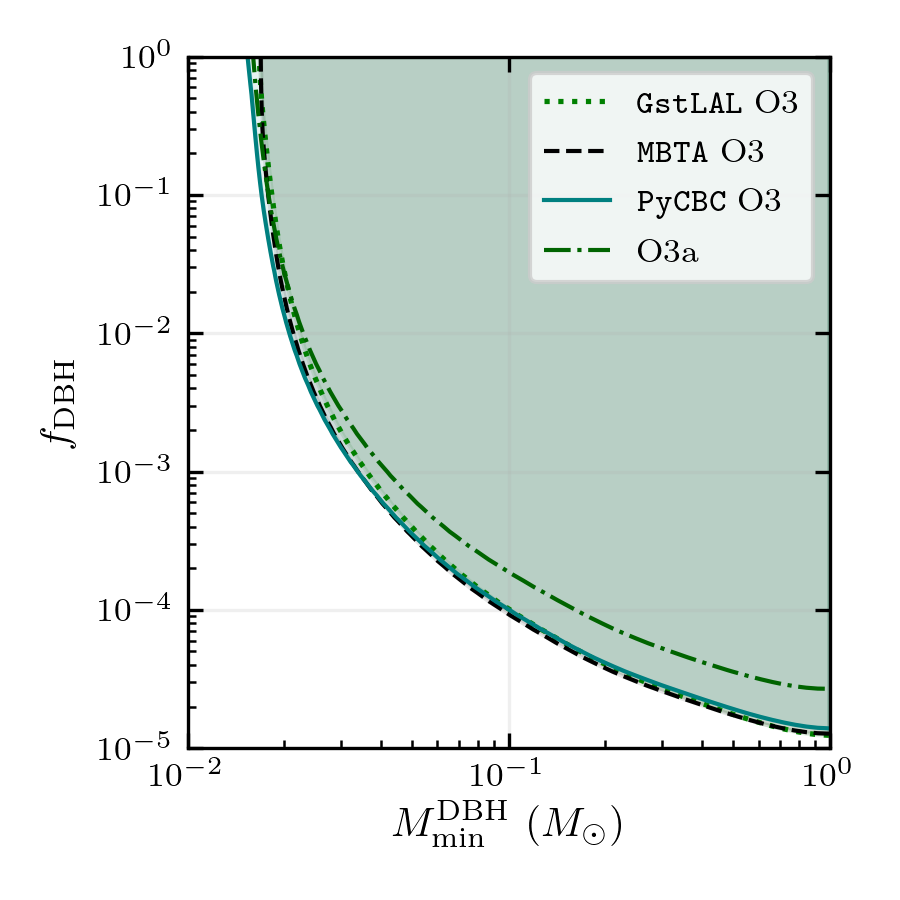}
\caption{\label{fig:dissipativeDM} Constraints on the abundance of DBHs, $f_\mathrm{DBH}$, as a function of the lower limit of the DBH mass distribution, $M_{\rm min}^{\rm DBH}$ from O3 data for the 3 search pipelines: \texttt{GstLAL} (\textit{dotted}), \texttt{MBTA} (\textit{dashed}) and \texttt{PyCBC} (\textit{solid}). Constraints from the search for \SSM{} compact objects in O3a data~\citep{LIGOScientific:2021job} are shown for comparison.}
\end{figure}

The upper limits on $f_{\rm DBH}$ are shown as a function of $M_{\rm min}^{\rm DBH}$ in Fig.~\ref{fig:dissipativeDM}. Compared to the results obtained from the \SSM{} search in O3a~\citep{LIGOScientific:2021job}, where the most stringent constraint on $f_{\rm DBH} \lesssim 0.003\% $, the limit improves by roughly a factor of 2, which can be directly attributed to the increase in the observing time. We derive the strictest limit on $f_{\rm DBH} \lesssim 0.0012--0.0014\% $ at $M_{\rm min}^{\rm DBH} =1 M_\odot$ {across the 3 pipelines}. The range of heavy dark fermion masses, $m_{\rm X}$ probed by this search inferred from the Chandrasekhar limit of the fermionic particle progenitors of DBHs, is  $1.1 \thinspace \mathrm{GeV}/c^2 < m_{\rm X} < 8.9 \thinspace \mathrm{GeV}/c^2$.

A non-detection provides no information for the model parameter $M_{\rm min}^{\rm DBH} < 2\times 10^{-2} M_\odot$ because the searches are not sensitive enough to support distributions with $M_{\rm min}^{\rm DBH}$ in that mass range since we only consider $M_{\rm max}^{\rm DBH} = rM_{\rm min}^{\rm DBH}$ with $2 \leq r \leq 1000$. We also exclude limits where $M_{\rm min}^{\rm DBH} > 1 M_\odot$ because the detection of a \SSM{} DBH would require a mass distribution with $M_{\rm min}^{\rm DBH} \leq 1 M_\odot$. If these limits survive with subsequent searches, the detection of a \SSM{} compact object would directly constrain the particle properties of atomic dark matter. Future searches could potentially rule out regions of the \DM{} parameter space associated with dissipative dark matter.

\section{Conclusions and outlook}\label{sec:conc}
We have presented a search for compact binary coalescences with at least one SSM component in data from the second half of the third LVK observing run, O3b. The search did not yield any significant candidates.

The absence of significant candidates enables us to set improved merger-rate limits based on the full O3 dataset. We obtain consistent results with each of the three considered search pipelines. We demonstrate how the new upper limits can be used to constrain two illustrative models: SSM PBHs and DBHs.

We have considered PBH merger rate models that incorporate additional physics relative to previous LVK works and obtained new limits that are less stringent than previous LVK searches for SSM objects. Using these upper limits, the data allow us to exclude
equal mass PBHs with a \DM{} fraction smaller than one, in the entire subsolar range probed by the search.  More general PBH distributions with extended mass functions remain viable, even for $f_{\rm PBH} \approx 1$. {Our SSM search therefore provides limits that are complementary to other types of observations such as pulsar timing arrays~\citep{DeLuca:2020agl,Chen:2019xse,Domenech:2020ers,Kohri:2020qqd} and microlensing surveys~\citep{Allsman:2000kg,Tisserand:2006zx,2011MNRAS.416.2949W}} that can probe or constrain the GW background induced by the density fluctuations at the origin of the formation of SSM PBHs.

For the dissipative dark matter model we consider, bounds on dark matter self-interactions on large scales \citep{Markevitch:2003at} already weakly constrain the amount of dark matter that can be efficiently cooling, so only some of the dark matter can have cooled sufficiently to form compact objects \citep{Buckley:2017ttd,Shandera:2018xkn}. Our analysis here provides the strongest constraint on this fraction so far from a \SSM{} search, finding that no more than $f_{\rm DBH}\approx10^{-5}$ of atomic dark matter can be collapsed into black holes for distributions that include DBHs in the $0.2$--$1M_{\odot}$ range where the sensitive volume is determined from this search alone.

Given the fundamental physics implications of observing a \SSM{} black hole, it will be important to continue this type of search in the next LVK observing runs~\citep{LVKarticle}.  
Each of the upcoming observing runs will be preceded by detector upgrades, designed to enhance the sensitivity of our ground-based interferometer network and our reach into the Universe. These developments will facilitate either the detection of a \SSM{} compact object or provide tighter constraints on their abundance.

\section*{Acknowledgments}

Analyses in this catalog relied upon the LALSuite software library~\citep{lalsuite-software}.
The detection of the signals and subsequent significance evaluations in this catalog were performed with the GstLAL-based inspiral software pipeline~\citep{Messick:2016aqy,Sachdev:2019vvd,Hanna:2019ezx,Cannon:2020qnf}, with the MBTA pipeline~\citep{Adams:2015ulm,Aubin:2020goo}, and with the PyCBC~\citep{Usman:2015kfa,Nitz:2017svb,Davies:2020tsx} package. 
Plots were prepared with Matplotlib~\citep{Hunter:2007ouj}. 
Numpy~\citep{Harris:2020xlr} and Scipy~\citep{Virtanen:2019joe} were used in the preparation of the manuscript.

This material is based upon work supported by NSF's LIGO Laboratory which is a major facility
fully funded by the National Science Foundation.
The authors also gratefully acknowledge the support of
the Science and Technology Facilities Council (STFC) of the
United Kingdom, the Max-Planck-Society (MPS), and the State of
Niedersachsen/Germany for support of the construction of Advanced LIGO 
and construction and operation of the GEO\,600 detector. 
Additional support for Advanced LIGO was provided by the Australian Research Council.
The authors gratefully acknowledge the Italian Istituto Nazionale di Fisica Nucleare (INFN),  
the French Centre National de la Recherche Scientifique (CNRS) and
the Netherlands Organization for Scientific Research (NWO), 
for the construction and operation of the Virgo detector
and the creation and support  of the EGO consortium. 
The authors also gratefully acknowledge research support from these agencies as well as by 
the Council of Scientific and Industrial Research of India, 
the Department of Science and Technology, India,
the Science \& Engineering Research Board (SERB), India,
the Ministry of Human Resource Development, India,
the Spanish Agencia Estatal de Investigaci\'on (AEI),
the Spanish Ministerio de Ciencia e Innovaci\'on and Ministerio de Universidades,
the Conselleria de Fons Europeus, Universitat i Cultura and the Direcci\'o General de Pol\'{\i}tica Universitaria i Recerca del Govern de les Illes Balears,
the Conselleria d'Innovaci\'o, Universitats, Ci\`encia i Societat Digital de la Generalitat Valenciana and
the CERCA Programme Generalitat de Catalunya, Spain,
the National Science Centre of Poland and the European Union --- European Regional Development Fund; Foundation for Polish Science (FNP),
the Swiss National Science Foundation (SNSF),
the Russian Foundation for Basic Research, 
the Russian Science Foundation,
the European Commission,
the European Social Funds (ESF),
the European Regional Development Funds (ERDF),
the Royal Society, 
the Scottish Funding Council, 
the Scottish Universities Physics Alliance, 
the Hungarian Scientific Research Fund (OTKA),
the French Lyon Institute of Origins (LIO),
the Belgian Fonds de la Recherche Scientifique (FRS-FNRS), 
Actions de Recherche Concert\'{e}es (ARC) and
Fonds Wetenschappelijk Onderzoek --- Vlaanderen (FWO), Belgium,
the Paris \^{I}le-de-France Region, 
the National Research, Development and Innovation Office Hungary (NKFIH), 
the National Research Foundation of Korea,
the Natural Science and Engineering Research Council Canada,
Canadian Foundation for Innovation (CFI),
the Brazilian Ministry of Science, Technology, and Innovations,
the International Center for Theoretical Physics South American Institute for Fundamental Research (ICTP-SAIFR), 
the Research Grants Council of Hong Kong,
the National Natural Science Foundation of China (NSFC),
the Leverhulme Trust, 
the Research Corporation, 
the Ministry of Science and Technology (MOST), Taiwan,
the United States Department of Energy,
and
the Kavli Foundation.
The authors gratefully acknowledge the support of the NSF, STFC, INFN and CNRS for provision of computational resources. Funding for this project was provided by the Charles E. Kaufman Foundation of The Pittsburgh Foundation and the Institute for Computational and Data Sciences at Penn State. 

This work was supported by MEXT, JSPS Leading-edge Research Infrastructure Program, JSPS Grant-in-Aid for Specially Promoted Research 26000005, JSPS Grant-in-Aid for Scientific Research on Innovative Areas 2905: JP17H06358, JP17H06361 and JP17H06364, JSPS Core-to-Core Program A. Advanced Research Networks, JSPS Grant-in-Aid for Scientific Research (S) 17H06133 and 20H05639 , JSPS Grant-in-Aid for Transformative Research Areas (A) 20A203: JP20H05854, the joint research program of the Institute for Cosmic Ray Research, University of Tokyo, National Research Foundation (NRF), Computing Infrastructure Project of KISTI-GSDC, Korea Astronomy and Space Science Institute (KASI), and Ministry of Science and ICT (MSIT) in Korea, Academia Sinica (AS), AS Grid Center (ASGC) and the Ministry of Science and Technology (MoST) in Taiwan under grants including AS-CDA-105-M06, Advanced Technology Center (ATC) of NAOJ, and Mechanical Engineering Center of KEK.

{\it We would like to thank all of the essential workers who put their health at risk during the COVID-19 pandemic, without whom we would not have been able to complete this work.}

\section*{Data availability} The raw data used in the analyses are available via the \href{https://www.gw-openscience.org/}{Gravitational Wave Open Science Center}. The derived data generated in this work can be found on the \href{https://dcc.ligo.org/LIGO-P2200139/public}{LIGO Document Control Center}. 

\clearpage % for now so that all figs stay in place

\bibliographystyle{mnras}
\bibliography{references}

\end{document}